\documentclass[epj]{svjour}
\usepackage{graphics}
\usepackage{graphicx}
\usepackage{enumerate}
\begin{document}
\title{Important Influence of Entrance Channel Reorientation Coupling on Proton Stripping} 
\author{N. Keeley\inst{1} \and K. W. Kemper\inst{2,3} \and K. Rusek\inst{3}
}                     % Do not remove
\institute{National Centre for Nuclear Research, ul.\ Andrzeja So\l tana 7, 05-400 Otwock, Poland
\and Department of Physics, Florida State University, Tallahassee, Florida 32306, USA
\and Heavy Ion Laboratory, University of Warsaw, ul.\ Pasteura 5a, 02-093 Warsaw, Poland}
\date{Received: date / Revised version: date}
% The correct dates will be entered by Springer
%
\abstract{While it is well established that the ground state reorientation coupling can have a
significant influence on the elastic scattering of deformed nuclei, the effect of such couplings
on transfer channels has been much less well investigated. In this letter we demonstrate 
that the $^{208}$Pb($^7$Li,$^6$He)$^{209}$Bi proton stripping reaction 
at an incident energy of 52 MeV can be well described by the inclusion of the $^7$Li ground state
reorientation coupling within the coupled channels Born approximation formalism. Full finite-range
distorted wave Born approximation calculations were previously found to be unable to describe these
data. Addition of coupling to the 0.478-MeV $1/2^-$ excited state of $^7$Li, together
with the associated two-step transfer path, has little or no influence on the shape of the angular
distributions (except that for stripping leading to the 1.61-MeV $13/2^+$ level of $^{209}$Bi
which is significantly improved) but does affect appreciably the values of the $^{209}\mathrm{Bi}
\rightarrow \protect{^{208}\mathrm{Pb}} + p$ spectroscopic factors. Implications
for experiments with weakly-bound light radioactive beams are discussed.
} %end of abstract
\authorrunning{N. Keeley {\em et al.\/}}
\titlerunning{Important Influence \ldots}
\maketitle
While the distorted wave Born approximation in its full finite-range version (FR-DWBA) has been
applied to the analysis of a wide body of heavy-ion reaction data with considerable success, it has been
known since the early 1970s that certain classes of reaction, in particular (but by no means restricted
to) single-proton transfers, cannot be satisfactorily described using it without recourse to such undesirable
expedients as {\it ad hoc} adjustments of the exit channel optical potential parameters. These adjustments
are usually such that the resulting parameters no longer describe the relevant elastic scattering, thus violating
one of the fundamental tenets of the DWBA. Many of these reactions are poorly matched, either in terms of
Q value or transferred angular momentum, and at the time this led to the conjecture that the inclusion of two-step
reaction paths via excited states of the projectile and/or ejectile within the coupled channels Born approximation
(CCBA) formalism would enable good descriptions to be obtained without the need arbitrarily to modify the
exit channel distorting potentials. However, the prohibitive computational overhead of such calculations
with then available resources precluded the testing of this hypothesis.  

There the question has remained in many cases for the intervening four decades as interest in heavy-ion
reactions waned. However, with the advent of radioactive beams of sufficient intensity and optical quality 
there has been a resurgence in this field and it seems timely to revisit some of the stable beam data
with a view to establishing whether they can, in fact, be satisfactorily described by including likely
two-step reaction paths. Current easy availability of significant computing power makes the routine 
application of the full arsenal of direct reaction theory to the interpretation of such data a practical
possibility. In a recent article \cite{Kee20} we showed that data \cite{Tot76,Oer84} for the $^{208}$Pb($^{12}$C,$^{11}$B)$^{209}$Bi
single-proton stripping reaction which the FR-DWBA failed to explain could be well described by CCBA calculations
including two-step transfer via the 4.44-MeV $2^+_1$ excited state of $^{12}$C using shell model spectroscopic
amplitudes for the projectile-like overlaps and distorting potentials that fitted the relevant elastic
scattering data in both entrance and exit channels. Similar calculations for the $^{208}$Pb($^{12}$C,$^{13}$C)$^{207}$Pb
single-neutron pickup improved the already good description of the corresponding data \cite{Tot76,Oer84} by the
FR-DWBA. 

Here we show that the 52-MeV $^{208}$Pb($^7$Li,$^6$He)$^{209}$Bi single-proton stripping
data of Zeller {\it et al.\/} \cite{Zel79}, which the FR-DWBA also failed to describe satisfactorily, may be
fitted simply by including the ground state reorientation coupling of the $^7$Li projectile, provided the   
description of the entrance channel elastic scattering data is maintained when the coupling is included. 
While the influence of ground state reorientation coupling on the elastic scattering and excited state
reorientation couplings on the inelastic scattering is well established (see, e.g., \cite{Hni81} and 
\cite{Vid76}), we are not aware of any previous studies that have demonstrated a significant effect of
entrance channel ground state reorientation coupling on {\em transfer} channels.
Addition of the excitation of the $1/2^-$ first excited state of $^7$Li and the two-step transfer path
via this state produces a barely perceptible improvement in the description of the stripping data, with
the exception of that for the 1.61-MeV $13/2^+$ level of $^{209}$Bi which exhibits a significant further
improvement when this coupling is included. We discuss some of the implications of these results for experiments
involving weakly-bound light radioactive beams.  

All calculations were carried out with the code {\sc fresco} \cite{Tho88}.
Inputs to the CCBA calculations were kept as similar as possible to those of the original FR-DWBA 
calculations of Ref.\ \cite{Zel79}, with the exception that the spectroscopic 
amplitudes for the $\left<^7\mathrm{Li} \mid \protect{^6\mathrm{He}} + p\right>$ overlaps were taken
from \cite{Rud05} rather than Cohen and Kurath \cite{Coh67} in order to have a consistent set for all
the overlaps required. However, the spectroscopic amplitudes of Refs.\ \cite{Rud05} and \cite{Coh67}
for the $^7\mathrm{Li}(3/2^-) \rightarrow \protect{^6\mathrm{He}(0^+)} + p$ transition only differ by about 4\%.
The $^6\mathrm{He} + p$ and $^{208}\mathrm{Pb} + p$ binding potentials were as in Ref.\ \cite{Zel79}.
We also retained the exit channel $^6\mathrm{He} + \protect{^{209}\mathrm{Bi}}$ optical potential of
Ref.\ \cite{Zel79}. While this was obtained from a fit to $^6$Li + $^{209}$Bi elastic scattering data
at the appropriate energy, the comparative study of Ref.\ \cite{Kuc09} suggests that for incident energies
this far above the Coulomb barrier $^6$He elastic scattering should be adequately described by $^6$Li
optical potential parameters, even for heavy targets like $^{209}$Bi. Coupled discretised continuum
channels (CDCC) calculations using the modified dineutron model of Moro {\it et al.\/} \cite{Mor07} confirmed
this, since the calculated elastic scattering angular distribution was well described by the exit channel 
potential parameters of Ref.\ \cite{Zel79}.

\begin{figure}
\includegraphics[width=\columnwidth,clip=]{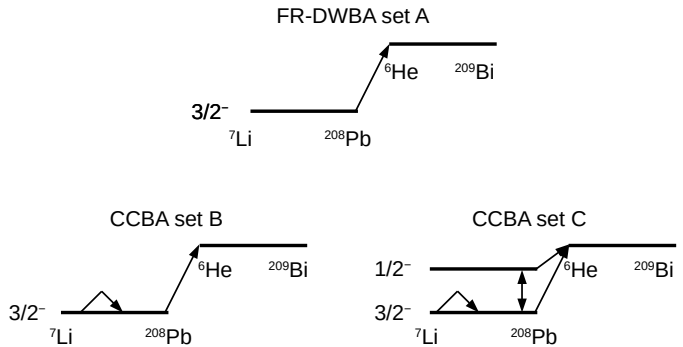}
\caption{\label{fig:coup} Schematic representations of the coupling schemes corresponding to FR-DWBA (set A), CCBA with
$^7$Li ground state reorientation coupling only (set B) and CCBA with $^7$Li ground state reorientation
plus coupling to the 0.478-MeV $1/2^-$ level (set C).
}
\end{figure}

The $^7$Li couplings in the entrance partition were included using standard rotational model form factors
with the nuclear multipoles calculated by numerically deforming the radii of the diagonal potential and
projecting by Gaussian quadrature onto the required multipoles. Use of the more approximate derivative
method to calculate the nuclear multipoles gave almost identical results provided that the nuclear
deformation length was re-adjusted as well as the optical potential parameters to give the same
inelastic and elastic scattering angular distributions as the calculations using the more accurate
Gaussian projection method. 
The $3/2^-$ ground state and 0.478-MeV $1/2^-$ first excited state were treated as members of a prolate
K = 1/2 band, with the Coulomb coupling strength $B(E2; 3/2^- \rightarrow 1/2^-) = 8.3 \pm 0.5$ e$^2$fm$^4$
taken from Ref.\ \cite{Wel85} and the nuclear deformation
length $\delta_2 = 2.0$ fm from Ref.\ \cite{Wei19}. Two CCBA calculations were carried out, the first
including the $^7$Li ground state reorientation coupling only and the second the ground state reorientation
plus excitation of the $1/2^-$ state together with the two-step transfer path via this
state. In both cases the optical potential parameters were
readjusted to recover the fit to the $^7$Li + $^{208}$Pb elastic scattering data of Ref.\ \cite{Zel79}
and the resulting values are listed in Table \ref{tab:omp} as sets B and C respectively, together with the
original parameters of Ref.\ \cite{Zel79} (set A) for ease of reference.
\begin{table}
\caption{Parameters of the $^7$Li + $^{208}$Pb optical model potentials used in the FR-DWBA (Set A) 
and CCBA calculations (sets B
and C). All radii use the convention: $R_x = r_x \times \mathrm{A_T}^{1/3}$ fm. The Coulomb radius 
parameter $r_\mathrm{C} = 1.40$ fm in all cases.}
\label{tab:omp} 
\begin{tabular}{lllllll}
\hline\noalign{\smallskip}
Set & $V$ & $r_V$ & $a_V$ & $W$ & $r_W$ & $a_W$ \\
 & (MeV) & (fm) & (fm) & (MeV) & (fm) & (fm) \\
\noalign{\smallskip}\hline\noalign{\smallskip}
A \cite{Zel79} & 293.8 & 1.253 & 0.785 & 18.99 & 1.602 & 0.743 \\
B & 321.7 & 1.253 &  0.785 &  14.65 & 1.602 & 0.802 \\
C & 352.0 & 1.253 & 0.785 & 17.88 & 1.602 & 0.759 \\
\noalign{\smallskip}\hline
\end{tabular}
\end{table}
The relevant coupling schemes are presented schematically in Fig.\ \ref{fig:coup}.
All calculations used the post form of the DWBA and included the full complex remnant term.

\begin{figure}
\includegraphics[width=\columnwidth,clip=]{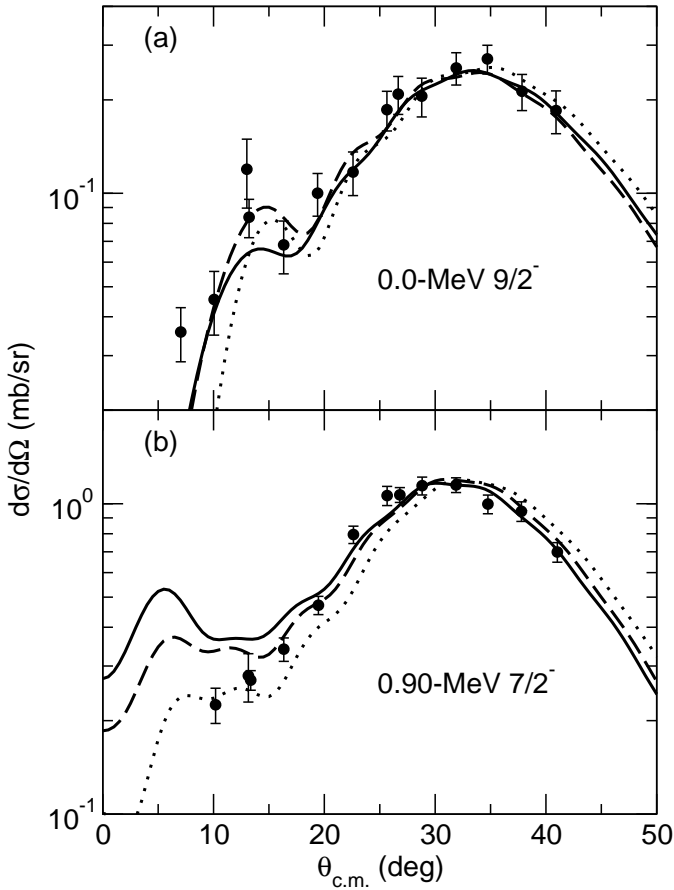}
\caption{\label{fig1} The $^{208}$Pb($^7$Li,$^6$He)$^{209}$Bi single-proton stripping data of Ref.\ \cite{Zel79} 
for transitions populating the 0.0-MeV $9/2^-$ (a) and 0.90-MeV $7/2^-$ (b) levels of $^{209}$Bi (solid circles) 
compared with the results of FR-DWBA calculations (dotted curves), CCBA calculations including the $^7$Li ground 
state reorientation coupling only (dashed curves) and ground state reorientation plus excitation of the 0.478-MeV 
$1/2^-$ state (solid curves). The latter also include the two-step transfer via the $^7$Li excited state. 
} 
\end{figure}

\begin{figure}
\includegraphics[width=\columnwidth,clip=]{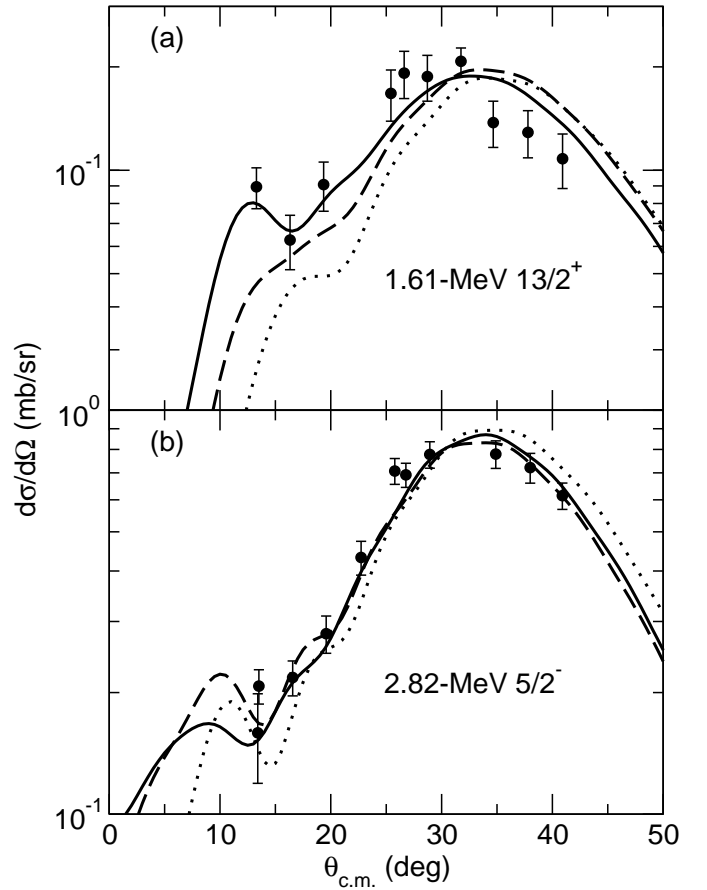}
\caption{\label{fig2} As for Fig.\ \ref{fig1} but for transitions populating the 1.61-MeV $13/2^+$ (a) and 
2.82-MeV $5/2^-$ (b) levels of $^{209}$Bi.
}
\end{figure}

\begin{figure}
\includegraphics[width=\columnwidth,clip=]{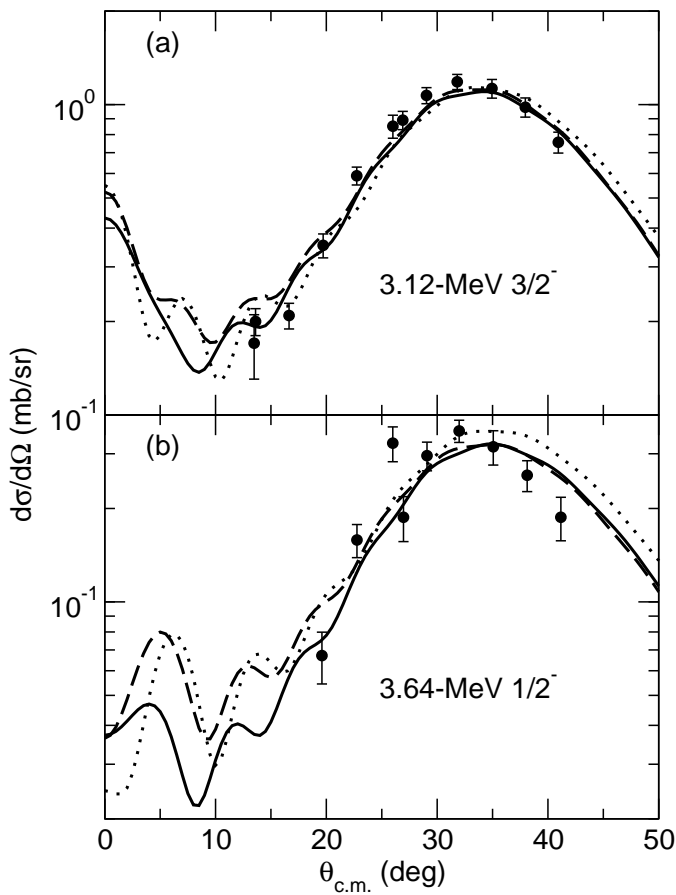}
\caption{\label{fig3} As for Fig.\ \ref{fig1} but for transitions populating the 3.12-MeV $3/2^-$ (a) and 
3.64-MeV $1/2^-$ (b) levels of $^{209}$Bi.
}
\end{figure}

In Figs.\ \ref{fig1} -- \ref{fig3} we compare the results of FR-DWBA calculations with the data of Ref.\ \cite{Zel79}.
While the data for stripping to the 0.0-MeV $9/2^-$ and 3.12-MeV $3/2^-$ levels are reasonably
well described the calculations for the other levels show significant shifts in the position of the peak to larger 
angles compared with the data. This is particularly striking in the case of the 1.61-MeV $13/2^+$ level, analysis
of which was not pursued in Ref.\ \cite{Zel79} due to the prohibitive computing time required with the then
available resources.

Figures \ref{fig1} -- \ref{fig3} also compare the stripping data with the results of CCBA calculations including the
$^7$Li ground state reorientation only (dashed curves) and ground state reorientation plus excitation of the
0.478-MeV $1/2^-$ excited state (solid curves), the latter also including the two-step transfer path via the 
$^7$Li excited state. Simply including the $^7$Li ground state reorientation coupling enables a good description
of all the data, with the exception of that for stripping to the $13/2^+$ level of $^{209}$Bi which, while
considerably improved, could be better. Addition of the coupling to the 0.478-MeV $1/2^-$ state and the associated
two-step transfer path further improves the description of this level, with a relatively minor influence 
on the description of the other levels. Note that the good description of the stripping
data by the CCBA calculations is conditional on the refitting of the entrance channel elastic scattering
data with the relevant $^7$Li couplings included. While the extraction of spectroscopic factors is not the goal of this
work, we also note that in addition to improving the description of the angular distributions the inclusion
of the ground state reorientation coupling alone and the reorientation plus inelastic excitation couplings
(together with the relevant two-step transfer path) both have a significant effect on the extracted values
of the spectroscopic factors for the $\left<^{209}\mathrm{Bi} \mid \protect{^{208}\mathrm{Pb}} + p\right>$
overlaps, see Table \ref{tab:sfs}. 
\begin{table*}
\caption{Spectroscopic factors (C$^2$S) for the $\left<^{209}\mathrm{Bi} \mid \protect{^{208}\mathrm{Pb}} + p\right>$
overlaps extracted from the FR-DWBA (set A) and CCBA calculations including the $^7$Li ground state reorientation
coupling only (set B) and ground state reorientation plus excitation of the 0.478-MeV $1/2^-$ excited
state together with the relevant two-step transfer path (set C).}
\label{tab:sfs}
\begin{center}
\begin{tabular}{lcccccc}
\hline\noalign{\smallskip}
Set & 0.0-MeV & 0.90-MeV & 1.61-MeV & 2.82-MeV & 3.12-MeV & 3.64-MeV \\
 & $9/2^-$ & $7/2^-$ & $13/2^+$ & $5/2^-$ & $3/2^-$ & $1/2^-$ \\
\noalign{\smallskip}\hline\noalign{\smallskip}
A & 1.40 & 1.19 & 0.81 & 1.07 & 0.89 & 0.64 \\
B & 1.51 & 1.31 & 0.88 & 1.15 & 1.09 & 0.73 \\
C & 1.68 & 1.55 & 0.98 & 1.39 & 1.34 & 0.90 \\
\noalign{\smallskip}\hline
\end{tabular}
\end{center}
\end{table*}

It has long been known that the ground state reorientation coupling has an important
influence on the analysing powers in reactions induced by polarised beams of $^7$Li, the elastic scattering
second-rank tensor analysing powers in particular being mainly due to this coupling. It has also been
demonstrated to have a significant effect on the elastic scattering differential cross section angular
distribution \cite{Hni81}. However, in this work we present a case where the reorientation coupling {\it alone} has
an important influence on a transfer channel; in previous work of this type the $^7$Li ground state reorientation
has usually been associated with coupling to the 0.478-MeV $1/2^-$ excited state, so that although significant
effects on transfer reaction angular distributions were observed, the contribution of the reorientation coupling
alone was not established. 

We conclude with a number of observations concerning the results of this study and their bearing on the
analysis of transfer reactions induced by radioactive ion beams. We have demonstrated, for possibly the first
time, a significant influence on the transfer reaction differential cross section angular distributions
calculated within the CCBA framework of the ground state reorientation coupling in a deformed projectile 
{\it alone}. Since many light radioactive nuclei have similar or larger quadrupole moments to 
that of $^7$Li ($Q = -40.6 \pm 0.8$ mb \cite{Die88}), in particular $^8$Li ($Q = +32.7 \pm 0.6$ mb \cite{Min93}), 
$^9$Li ($Q = -27.4 \pm 1.0$ mb \cite{Arn92}) and $^{11}$Li ($Q = -33.3 \pm 0.5$ mb \cite{Neu08}) but especially $^8$B
($Q = 68.3 \pm 2.1$ mb \cite{Min92,Min93a}), this suggests that inclusion of the ground state reorientation 
coupling in the analysis of transfer reactions involving such nuclei is advisable. Also, the need to ensure
that the elastic scattering data remained well described by the calculations including the reorientation and
inelastic couplings in order to be able to describe the proton stripping data, underlines the importance of
measuring at least the entrance channel elastic scattering as well as the transfer process(es) of
interest. This can be particularly important in radioactive beam work where, owing to the peculiarities
of individual nuclei of this type, the use of ``near enough'' optical potential parameters may be misleading.

It is also clear from Table \ref{tab:sfs} that inclusion of both ground state reorientation and excitation
of the 0.478-MeV $1/2^-$ state of $^7$Li has a significant effect on the $^{209}\mathrm{Bi} \rightarrow
\protect{^{208}\mathrm{Pb}} + p$ spectroscopic factors extracted from the fits to the data. We particularly
wish to note that while the inclusion of the excitation of the $1/2^-$ state has little or no effect on the
shapes of the transfer angular distributions---with the exception of that for stripping populating the 1.61-MeV
$13/2^+$ level of $^{209}\mathrm{Bi}$---its influence on the extracted spectroscopic factors is far from
negligible. Thus, even though including the ground state reorientation coupling alone provides a good description
of the stripping data, if the extraction of spectroscopic factors is the aim of the analysis stopping at this
point could lead to erroneous conclusions. Indeed, it may well be that the addition of further couplings, e.g.\
an explicit treatment of the $^7\mathrm{Li} \rightarrow \alpha + t$ breakup via the CDCC method, will
be required to obtain a ``converged'' set of values for the $^{209}\mathrm{Bi} \rightarrow \protect{^{208}\mathrm{Pb}} 
+ p$ spectroscopic factors using this reaction. This illustrates one aspect of a question raised by Satchler in
the introduction to his book {\it Direct Nuclear Reactions} \cite{Sat83}: ``At this point one may ask,
`Where does one stop?' \ldots If the shape (angular distribution) of the [DWBA cross section] agrees with
experiment, it is trivial to extract a structure factor from the magnitude needed to match the data. This is
no longer true when multistep processes are studied; the structure amplitudes (and their signs) are involved in
an intimate way.'' 

In summary, we have shown that including the $^7$Li ground state reorientation coupling enabled a good
description of the $^{208}$Pb($^7$Li,$^6$He)$^{209}$Bi single-proton stripping data of Ref.\
\cite{Zel79} which FR-DWBA calculations were unable to fit. Addition of coupling to the 0.478-MeV $1/2^-$
state of $^7$Li plus the corresponding two-step transfer path had only a minor effect on the shapes of the
angular distributions, except for that leading to the 1.61-MeV $13/2^+$ level of $^{209}$Bi, but did have
a significant impact on the $^{209}\mathrm{Bi} \rightarrow \protect{^{208}\mathrm{Pb}} + p$ spectroscopic factors.
A good description of the elastic scattering was necessary to obtain the good fit to the stripping data.   
Since many weakly-bound light radioactive nuclei have similar properties, these results strongly suggest that
such couplings should be included in analyses of transfer data obtained with beams of these nuclei if
erroneous conclusions are not to be drawn. They also underline the need to obtain as complete a data set as
possible---ideally elastic and inelastic scattering as well as the transfer process(es) of interest---to
facilitate the correct inclusion of such couplings.

\end{document}